\documentstyle[prd,preprint,aps,epsf]{revtex}  
              \def\d{\delta}
    \def\f{\phi}       
   \def\G{\Gamma}        \def\l{\lambda}
  \def\m{\mu}

\def\vf{\varphi}             

\def\CD{{\cal D}}   
   \def\CL{{\cal L}}

\def\ha{{1\over2}}
\newcommand{\be}{\begin{equation}}
\newcommand{\bea}{\begin{eqnarray}}
\newcommand{\ee}{\end{equation}}
\newcommand{\eea}{\end{eqnarray}}
\newcommand{\ba}[1]{\left(\begin{array}{#1}}
\newcommand{\ea}{\end{array}\right)}
\newcommand{\fr}[2]{\frac{#1}{#2}}
\newcommand{\nn}{\nonumber}

\def\tr{\hbox{Tr}\,}
\def\tildef{\tilde{\varphi}}

\def\prd#1#2#3{Phys. Rev. D {\bf {#1}}, {#2} ({#3})}
\def\prs#1#2#3{Phys. Rev. {\bf {#1}}, {#2} ({#3})}

\def\plb#1#2#3{Phys. Lett. {\bf B{#1}}, {#2} ({#3})}

\def\npb#1#2#3{Nucl. Phys. {\bf B{#1}}, {#2} ({#3})}

\def\ap#1#2#3{Ann. Phys. {\bf {#1}}, {#2} ({#3})}
\def\jkps#1#2#3{J. Kor. Phys. Soc. {\bf {#1}}, {#2} ({#3})}
\def\zp#1#2#3{Z. Phys. {\bf {#1}}, {#2} ({#3})}
\begin{document}                
\draft
\title{Perturbative Expansion around the Gaussian Effective Action: 
The Background Field Method}
\author{Geon Hyoung Lee%
        \footnote{email: {\sl ghlee@theory.yonsei.ac.kr}} 
and
        Jae Hyung Yee%
        \footnote{email: {\sl jhyee@phya.yonsei.ac.kr}}}
\address{Department of Physics and Institute for Mathematical Sciences\\
     Yonsei University\\
     Seoul 120-749, Korea}

\date{\today}
\maketitle
\begin{abstract}                
We develop a systematic method of the perturbative expansion around 
the Gaussian effective action based on the background field method.
We show, by applying the method to the quantum mechanical
anharmonic oscillator problem,
that even the first non-trivial correction terms greatly improve
the Gaussian approximation.
\end{abstract}
\newpage
\section{Introduction}   
The variational Gaussian approximation method has provided 
a convenient and practical device in obtaining non-perturbative 
information from various quantum field theories, 
and has played an important role in enlarging our knowledge on the 
non-perturbative nature of the theories\cite{barns}.
The main drawback of this method, however, is that it does not provide 
a systematic procedure to compute the correction terms
to the approximation.
One natural way to improve the Gaussian approximation is to use more 
general non-Gaussian trial wave functional 
for the variational approximation\cite{polley}.
Although this approach can easily be applied to the case of
quantum mechanics, 
it has not been very successful in the case of quantum field theory.

The efforts to establish the systematic approximation method based on the 
variational techniques have been made by authors of ref.\cite{cea} and 
the so-called optimized expansion method and variational perturbation 
theory have been formulated.
It is based on the fact that the Gaussian approximation is to approximate
an interacting theory by a free field theory with a new mass, that is 
determined by variational gap equation.
One constructs a complete set of energy eigenstates for this free theory and 
use it as a basis for perturbative expansion of the interacting theory.
Cea and Tedesco \cite{cea} used this variational perturbation theory to 
evaluate the second order correction terms to
the Gaussian effective potential 
for the (1+1) and (2+1) dimensional $\f^4$ theories.
This perturbation theory is conceptually simple and provides a very 
convenient systematic procedure to evaluate the correction terms to
the Gaussian effective potentials for time-independent systems.
For time-dependent systems such as those treated by Eboli, 
Jackiw and Pi \cite{barns}, however, one needs a method that can deal 
with more general effective action.

It is the purpose of this paper to develop a systematic method of computing
correction terms to the Gaussian effective action based on the background
field method(BFM)\cite{cklee}.
The background field method provides a very effective and simple way of 
computing effective action in the loop-expansion.
The reason why the procedure of computing effective action in BFM
becomes simplified is that one has rearranged the summation of diagrams
in such a way that the propagator
used in the Feynman rule already contains the part of the 1-loop
interaction effect.
We proceed one step further so that the zeroth order term in the 
expansion would give the Gaussian approximation, and the propagator used 
in  the expansion would be the one obtained in the Gaussian approximation.
This provides a very convenient and
simple method in computing order by order
correction terms to the Gaussian approximation of the effective action.

In the next section we give a brief review of BFM and develope the 
perturbative expansion method around the Gaussian effective action by using 
scalar $\f^4$ theory.
In section \ref{sec3}, we illustrate the effectiveness of the method 
by evaluating the first non-trivial correction terms to
the Gaussian effective 
potential of the (0+1)-dimensional $\f^4$ theory (anharmonic oscillator) 
and compare the result 
with those of various other methods.
We conclude with some discussions in the last section.

\section{Perturbative Expansion around the Gaussian Effective Action}
For a system described by the classical action,
\be
S[\f]=\int d^4x \CL[\f(x),\partial_\m\f(x)],
\ee
the generating functional for Green's functions is defined by 
the path integral, 
\be
<0|0>^J \equiv Z[J] \equiv e^{i W[J]} 
\equiv \int \CD \f e^{i S[\f]+iJ\f},
\ee
where $J(x)$ is the external source,
$W[J]$ is the generating functional for connected Green's functions,
and the integral convention,
$J \f \equiv \int d^n x J(x) \f(x)$ is used in the exponent.
The Green's functions can be obtained by
\be
<0_+|T[\f(x_1) \cdots \f(x_n)]|0_->^J=(\frac{\d}{i \d J(x_1)}) \cdots
(\frac{\d}{i \d J(x_n)}) e^{iW[J]}.
\ee
The vacuum expectation value of the field operator
in the presence of external source is defined by 
\be
\vf \equiv <\f(x)>^J = \frac{\d}{\d J(x)} W[J].
\ee
The effective action is then defined by the Legendre transformation;
\be
\G[\vf] \equiv W[J]-\vf J .
\ee
Taking the derivative of $\G[\vf]$ with respect to the external source $J$ 
gives
\be
\fr{\d}{\d \vf} \G[\vf] = -J,
\ee
which is of the same form as the classical equation of motion.

We now introduce a new action $S[\f+B]$ which is obtained by
shifting the field $\f$ by a background field $B$.
This new action gives a new generating functional for Green's functions
in the presence of the background field $B$,
\bea
\tilde{Z}[J,B] & \equiv & e^{i \tilde{W}[J,B]} \nn \\
 & \equiv & \int \CD \f e^{i S[\f+B]+iJ\f},
\label{ztilde}
\eea
and the vacuum expectation value of field operator is now defined by 
\be
\tildef \equiv \frac{\d}{\d J} \tilde{W}[J,B]
= \frac{\d}{i \d J} \log (\tilde{Z}[J,B]).
\label{tad}
\ee
Since Eq.\ (\ref{ztilde}) can be written as
\be
\tilde{Z}[J,B]=\int \CD \f e^{iS[\f]+iJ(\f-B)} \\
=Z[J] e^{-iJB},
\ee
the new generating functional $\tilde{W}[J,B]$ for connected Green's
functions
in the presence of the background field is related to the old one by,
\be
\tilde{W}[J,B] = W[J]-JB .
\label{pre}
\ee
By taking the derivative of Eq.\ (\ref{pre}) with respect to $J$,
one obtains the relation between $\tildef$ and $\vf$;
\be
\tildef = \vf-B.
\ee
The new effective action $\tilde{\G}[\tildef,B]$ in
the presence of the background field $B$ can then be written as
\bea
\tilde{\G}[\tildef,B] &\equiv& \tilde{W}[J,B]-J \tildef \nn \\
&=&(W[J]-JB)-J(\vf-B)  \nn  \\
&=&\G[\tildef+B].
\label{smain}
\eea
If we set $\tildef=0$, then Eq.\ (\ref{smain}) becomes
\be
\G[B]=\tilde{\G}[\tildef=0,B],
\ee
which implies that the effective action $\G[B]$ consists of
one $\tildef$-particle
irreducible diagrams with no external $\tildef$-lines.
Due to Eq.\ (\ref{tad}) and (\ref{smain}),
the effective action can be considered to be the 
extremum value of $\tilde{W}[J,B]$ with respect to the variation of the 
external source $J$.

We now consider the n-dimensional scalar $\f^4$ theory described
by the Lagrangian density,
\be
\CL = -\ha \f (i\CD^{-1})\f-\fr{\l}{4!} \f^4,
\ee
where $\CD^{-1}$ is defined by
\be
i \CD^{-1} \equiv (\partial^2+\m^2).
\ee

The generating functional for Green's functions in the presence of the 
background field $B$ is 
given by
\bea
\tilde{Z}[J,B] & \equiv & \int \CD \f e^{iS(\f+B)+i\f J} \nn \\
&=& det \sqrt{-K} \exp[\ha B \CD^{-1} B-\fr{i \l}{4!}B^4] 
\label{deri}   \\
& & \times 
\exp[(B \CD^{-1} -\fr{i \l}{3!} B^3) \fr{\d}{i \d J}] 
\exp [-\fr{i \l}{4!} (\fr{\d}{i\d J})^4 - \fr{i \l}{3!} B (\fr{\d}{i\d J})^3]  
 e^{\ha JKJ}, \nn
\eea
where $K$ is defined by
\be
K_{xy}^{-1} \equiv \CD_{xy}^{-1}-\fr{i \l}{2}B_x^2 \d_{xy}.
\label{pre2}
\ee
The effective action is given by 
\be
e^{i \G[B]} = \tilde{Z}[J,B] |_{\tildef=0}.
\ee
Note that the first line of Eq.\ (\ref{deri}) gives rise to the 
1-loop effective action and the last line, upon setting $\tildef=0$,
represents the higher loop contributions to the effective action.
This higher loop contributions can be evaluated
by using the Feynman rule with
the propagator given by $K_{xy}$ of Eq.\ (\ref{pre2}).
Thus the effective action consists of  one $\tildef$-particle 
irreducible bubble diagrams with no external lines, and the procedure of 
evaluating the higher loop contributions is simplified compared
to the conventional method.
The reason why it is simplified is that the propagators used in the Feynman
rule, Eq.\ (\ref{pre2}), already contains the interaction effect 
through the background field $B$, and 
the generating functional has been rearranged in such a way that 
the zeroth order term in the expansion, the first line of Eq.\ (\ref{deri}),
is the one-loop effective action

\begin{figure}
\centering
\vspace{2cm}
\mbox{\epsfxsize=400pt\epsffile{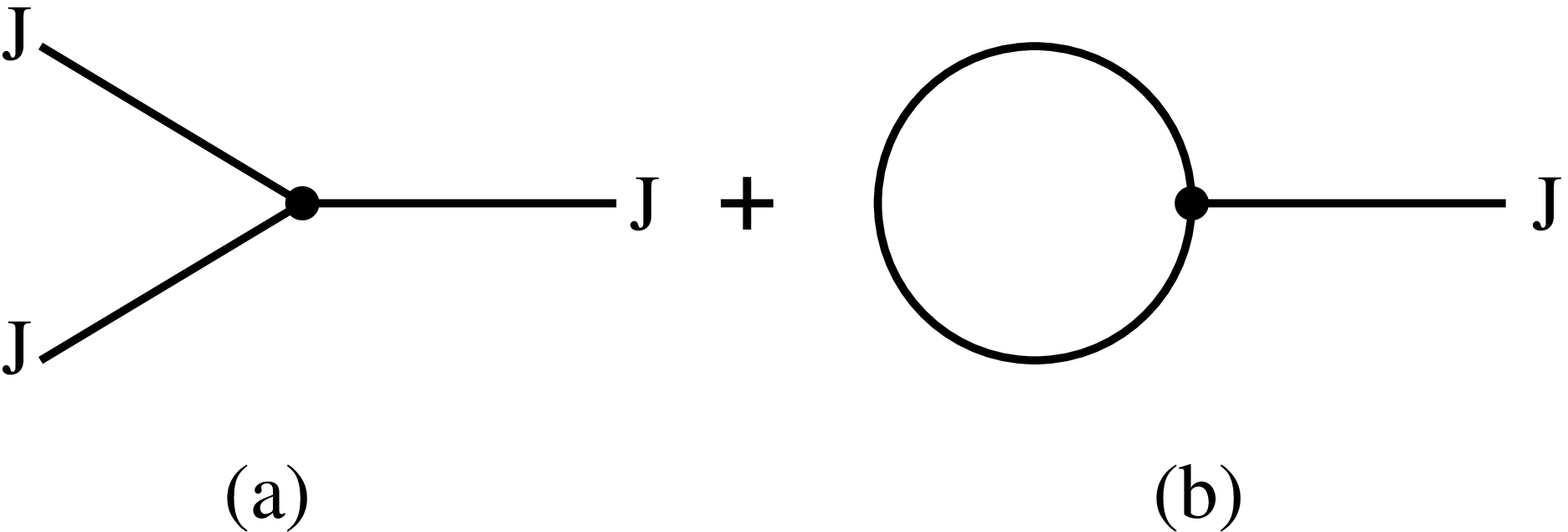}}
	\caption{Feynman diagrams for Eq.\ ({\protect \ref{j3}})}
\label{fig.j3}
\end{figure}

\begin{figure}
\centering
\vspace{2cm}
\mbox{\epsfxsize=400pt\epsffile{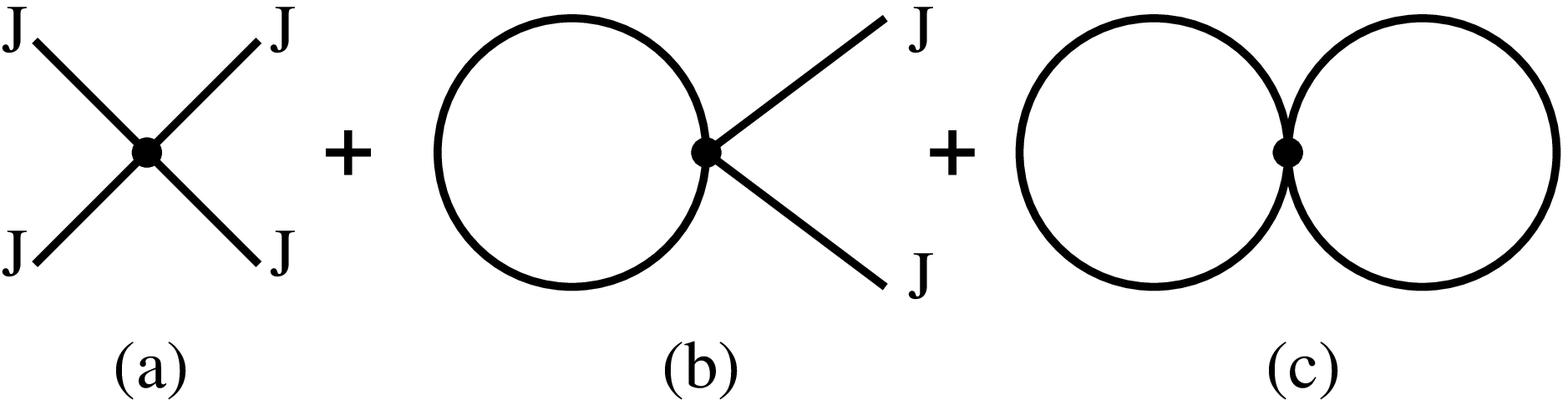}}
	\caption{Feynman diagrams for Eq.\ ({\protect \ref{j4}})}
\label{fig.j4}
\end{figure}

We now proceed one step further so that the propagator used
for the perturbative 
expansion would be the one obtained in the Gaussian approximation.
In other words, we want to rearrange $e^{i \G[B]}$ in such a way
that the zeroth order term
in the expansion would give the Gaussian approximation of
the effective action.
To do this, we consider the relations:
\bea
\fr{\d}{\d J_x} e^{\ha J_y G_{yz} J_z} &=&
G_{xy} J_y e^{\ha J_y G_{yz} J_z} \equiv
(GJ)_x e^{\ha JGJ} \\
\fr{\d^2}{\d J_x^2} e^{\ha J_y G_{yz} J_z} &=&
((GJ)_x^2+G_{xx}) e^{\ha JGJ} \\
\fr{\d^3}{\d J_x^3} e^{\ha J_y G_{yz} J_z} &=&
((GJ)_x^3+3G_{xx} (GJ)_x) e^{\ha JGJ}
\label{j3} \\
\fr{\d^4}{\d J_x^4} e^{\ha J_y G_{yz} J_z} &=&
((GJ)_x^4+6G_{xx} (GJ)_x^2+3G_{xx}^2) e^{\ha JGJ} ,
\label{j4}
\eea
which would appear in the expansion of the last line of Eq.\ (\ref{deri}).
Note that Eq.\ (\ref{j3}) and (\ref{j4}) contain the diagrams
where internal lines come
out of  a point and go back to the same point 
as shown in Fig. \ref{fig.j3} and \ref{fig.j4},
and the Gaussian approximation of the effective action
consists of such diagrams.

To extract such diagrams out of the last line of Eq.\ (\ref{deri}), 
we define new functional derivatives in such a way that
\bea
(\fr{\d^3}{\d J_x^3})' e^{\ha JGJ} &=&  ((GJ)_x)^3 e^{\ha JGJ}
\label{jp3}\\
(\fr{\d^4}{\d J_x^4})'  e^{\ha JGJ} &=& ((GJ)_x)^4 e^{\ha JGJ}.
\label{jp4}
\eea
Comparing Eqs.(\ref{j3}) and (\ref{j4}) and Eqs.(\ref{jp3}) and (\ref{jp4}),
we find that the primed derivatives must be defined by
\bea
(\fr{\d^3}{\d J_x^3})' & \equiv &
\fr{\d^3}{\d J_x^3}-3 G_{xx} \fr{\d}{\d J_x} 
\label{last3} \\
(\fr{\d^4}{\d J_x^4})' & \equiv &
\fr{\d^4}{\d J_x^3}-6 G_{xx} \fr{\d^2}{\d J_x^2}+3G_{xx}^2.
\label{last4}
\eea
We note that the primed derivatives operated on $e^{\ha JGJ}$ do not generate
the diagrams that contribute to the Gaussian approximation, such as those 
shown in Fig.\ref{fig.j3}(b), Fig.\ref{fig.j4}(b) and (c).

This implies that by using the primed derivatives one may extract the
diagrams that contribute to the Gaussian effective action,
out of the last line of Eq.\ (\ref{deri}).
To do this we need to find a Green's function $G$ which satisfies,
\bea
\lefteqn{\exp[\fr{\l B}{3!}(\fr{\d}{\d J})^3
-\fr{i \l}{4!}(\fr{\d}{\d J})^4] e^{\ha JKJ}= }   \nn   \\
& & N \exp[A \fr{\d}{\d J}] \exp[\fr{\l B}{3!}(\fr{\d^3}{\d J^3})'
  -\fr{i\l}{4!}(\fr{\d^4}{\d J^4})'] e^{\ha JGJ},
\label{dool}
\eea
where a normalization constant N and a function $A(x)$ are to be determined.
By using the definitions Eq.\ (\ref{last3}) and (\ref{last4}), 
one easily find $G$, $N$ and $A$ that satisfy
Eq.\ (\ref{dool}):
\bea
N &=&  \fr{det \sqrt{-G}}{det \sqrt{-(G^{-1}-\fr{i \l}{2} G_{xx})^{-1}}}
\exp [\fr{i\l}{8}G_{xx}^2],  \\
A &=& \fr{\l}{2} B G_{xx}, \\
G_{xy}^{-1} &=& K_{xy}^{-1}+\fr{i\l}{2}G_{xx} \d_{xy}  \nn \\
&=& \CD_{xy}^{-1}-\fr{i \l}{2}B_x^2 \d_{xy}+\fr{i\l}{2}G_{xx} \d_{xy}.
\label{pregap}
\eea
Thus we can rewrite the generating functional for Green's functions, 
Eq.\ (\ref{deri}) as
\bea
\lefteqn{\tilde{Z}[J,B]=\det\sqrt{-G} \exp[\ha B\CD^{-1}B
-\fr{i\l}{4!}B^4 +\fr{i\l}{8}G_{xx}^2]}  \nn \\ 
&\times \exp \left[ (-iB \CD^{-1}-\fr{\l}{3!}B^3 
+\fr{\l}{2} B G_{xx}) \fr{\d}{\d J} \right]
\exp[\fr{\l B}{3!}(\fr{\d^3}{\d J_x^3})'
  -\fr{i\l}{4!} (\fr{\d^4}{\d J_x^4})']
e^{\ha JGJ},
\eea
where the functional derivatives in the last line do not generate the cactus
type diagrams such as those shown in Fig.\ref{fig.j4}(c).

The effective action $\G[B]$ is given by 
\bea
e^{i \G[B]} &=& \tilde{Z}[J,B] |_{\tildef=0} \nn \\
&=& \det\sqrt{-G} \exp[\ha B\CD^{-1}B
-\fr{i\l}{4!}B^4 +\fr{i\l}{8}G_{xx}^2] \quad I[B],
\label{effec}
\eea
where
\bea
I[B] &=& \exp \left[ (-iB \CD^{-1}-\fr{\l}{3!}B^3 
+\fr{\l}{2} B G_{xx}) \fr{\d}{\d J} \right]  \nn \\
& &\times \left. \exp[\fr{\l B}{3!}(\fr{\d^3}{\d J_x^3})'
  -\fr{i\l}{4!} (\fr{\d^4}{\d J_x^4})'] e^{\ha JGJ} \right|_{\tildef=0}.
\label{ib}
\eea
Note that Eq.\ (\ref{pregap}), which determines the Green's function $G_{xy}$,
is exactly the variational gap equation written
in Minkowski space notation\cite{barns},
and therefore the coefficient of $I[B]$ in Eq.\ (\ref{effec}) gives rise to the 
Gaussian effective
action as we have alluded to above.
Since $I[B]$ of Eq.\ (\ref{ib}) can be expanded as a power series in $\l$
as is done in the conventional BFM,
we have the perturbative expansion of the effective action
around the Gaussian approximation.

\begin{figure}
\centering
\vspace{2cm}
\mbox{\epsfxsize=300pt\epsffile{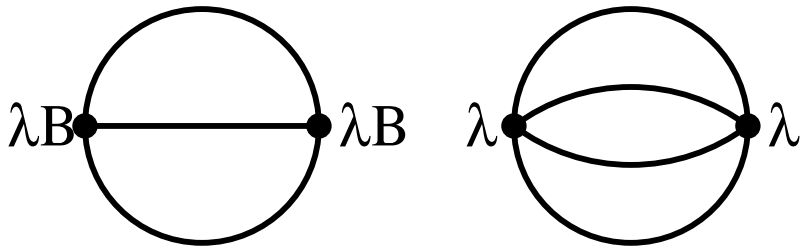}}
	\caption{Digrams contributing to the $\l^2$-order corrections to the
                    Gaussian effective action
}
\label{fig.second}
\end{figure}

We now consider the contributions coming from $I[B]$.
The  linear term in the exponent of Eq.\ (\ref{ib}) generate tadpole diagrams.
This term does not contribute to the effective action 
because the effective action $\G[B]$ is
the extremum value of $\tilde{W}[J,B]$
as explained earlier, and the tadpole term does not change
the extremum value.
We therefore see that $I[B]$ has the same structure as the last line
of Eq.\ (\ref{deri})
except that the propagator is replaced by $G_{xy}$,
and the functional derivatives are
replaced by the primed derivatives.
Thus $I[B]$ consists of one $\tildef$-particle irreducible
bubble diagrams with no 
external $\tildef$-lines,
and without the cactus type diagrams (with internal 
lines coming out of a point and going back to the same point).
We have thus established the systematic correction method to the
Gaussian approximation,
where one can compute order by order correction terms
to the Gaussian effective action.
One can easily show that the $\l^2$ contribution of $I[B]$ consists
of the diagrams 
shown in Fig.\ref{fig.second}.

If we consider the space-time independent background field
in the above formulation,
we obtain the effective potential defined by
\be
V_{eff}[B] \equiv -\fr{\G[B]}{\int d^nx}.
\ee
Since $B$ is a constant, $G_{xy}$ is a function of $(x-y)$.
Defining the Fourier transformation of $G_{xy}$ as 
\be
G_{xy} \equiv \int_p g(p) e^{ip(x-y)}
\label{furi}
\ee
and the effective mass as
\be
m^2 \equiv \m^2+\fr{\l}{2}B^2-\fr{\l}{2} \int_p g(p),
\label{mgis}
\ee
we can write Eq.\ (\ref{pregap}) as
\be
g(p)=\fr{1}{i(p^2-m^2)},
\label{gpis}
\ee
where
\be
\int_p \equiv \int \fr{d^np}{(2 \pi)^n},
\ee
and $n$ is the dimension of space-time.
We note that Eqs.(\ref{mgis}) and (\ref{gpis}) are the well known
mass gap equation for the Gaussian effective potential\cite{barns}.
Thus the zeroth order contribution of Eq.\ (\ref{effec})
is the Gaussian effective potential and 
the first non-trivial correction terms to the Gaussian approximation 
are of the order $\l^2$ as shown in Fig.\ref{fig.second}.

Thus up to the $\l^2$ order contribution of $I[B]$ the
effective potential is given by
\be
V_{eff} = V_G+V_P, 
\ee
where
\bea
V_G & \equiv & \fr{\m^2}{2} B^2+\fr{\l}{4!}B^4
-\fr{1}{2\l}(m^2-\m^2-\fr{\l}{2}B^2)^2
+i \tr \log \sqrt{-G},
\label{vgis} \\
V_P & \equiv & i \fr{ \l^2}{12}B^2 G^3 -i \fr{\l^2}{48} G^4,
\label{vpis}
\eea
and $G$ is given by Eqs.(\ref{furi}) $\sim$ (\ref{gpis}).
$V_G$ is the Gaussian effective potential, and $V_P$ is the first non-trivial 
contribution from $I[B]$ which is shown in Fig.\ref{fig.second}.
This result is the same as that of Cea and Tedesco\cite{cea} computed by
using the variational perturbation theory.

\section{Effective potential
for the (0+1)-dimensional $\f^4$ theory}
\label{sec3}
To illustrate how much our method improve the Gaussian approximation
we now consider the (0+1)-dimensional $\f^4$ theory,
which is the quantum mechanical 
anharmonic oscillator, 
and compare the results with those of various other methods.
By applying the Wick rotation, Eqs.(\ref{vgis}) and (\ref{vpis})
can be written in
this case as
\be
V_{eff}[B] = V_G+V_P ,
\label{pot.gen} 
\ee
where
\bea
V_G &=& \fr{\m^2}{2} B^2+\fr{\l}{24}B^4 
-\fr{1}{2\l}(m^2-\m^2-\fr{\l}{2}B^2)^2+ \hbar \fr{\sqrt{m^2}}{2},
\label{pot.gauss} \\
V_P &=& -\fr{\l^2 \hbar^2}{144 m^4} B^2 -
\fr{\l^2 \hbar^3}{1536 (\sqrt{m^2})^5}.
\label{pot.pert}
\eea
The effective mass, $m$, in Eqs.(\ref{pot.gauss})
and (\ref{pot.pert}) is given by
\be
m^2 = \m^2+\fr{\l}{2} B^2 + \fr{\l \hbar}{4 \sqrt{m^2}},
\label{msgap}
\ee
where we have put the Plank constant explicitly.

The improved Gaussian effective potential contains all
diagrams up to 2-loop.
Therefore, The 2-loop effective potential, which can be obtained by 
taking the terms up to $\hbar^2$ order from Eq.\ (\ref{pot.gen}), is
\bea
V_{2-loop} &=& \fr{\m^2}{2} B^2 + \fr{\l}{24} B^4 
+ \fr{\hbar }{2} \sqrt{\m^2+\fr{\l}{2} B^2}  \nn  \\
& & + \hbar^2 \left( \fr{\l}{32}(\m^2+\fr{\l}{2} B^2)^{-1}
-\fr{\l^2}{144}B^2 (\m^2+\fr{\l}{2} B^2)^{-2}   \right).
\label{loop2}
\eea
Ignoring the $\hbar^2$ term, we obtain the 1-loop effective potential,
which agrees with
that of Ref.\cite{steve1}.
Using these results, we evaluate the ground state energies
of various approximation methods and
compare their results.

For simplicity, we consider the case of positive $\m^2=1$,
and compare the ground state energies of 
various approximation methods in the unit $\hbar=1$.
Then the classical potential has minimum $V_{cl}=0$ at $B=0$.
The values of the coupling constant will be chosen to be 
that of \cite{steve1}.
($\l$ of \cite{steve1} corresponds to our $\fr{\l}{4!}$.)

Table \ref{ground} shows the ground state energy values of various
approximations
and their errors compared to the numerical 
results, for several coupling constants.
This shows that the first non-trivial correction
to the Gaussian approximation
greatly improves the result for the small coupling region.
For strong coupling, Gaussian and improved Gaussian
approximation display much 
better results than the loop expansion, 
which shows the non-perturbative nature of the
approximation methods.
Though the GEP is accurate to within about $2 \%$
in the strong coupling region, 
the error of the improved GEP is smaller than about $0.8 \%$.
This shows that the perturbative expansion method greatly
improves the Gaussian 
approximation even at the first non-trivial correction level.

\section{Discussion}
\label{chap.result}
We have developed the perturbative expansion method around the Gaussian
effective action by using the background field method.
The method is based on the observation that the Gaussian effective
action consists of cactus type diagrams,
which were extracted out of the functional derivative part of the effective
action, i.e., the last line of Eq.\ (\ref{deri}), by introducing the primed 
derivatives defined in Eqs.(\ref{last3}) and (\ref{last4}).
This procedure effectively rearranged the diagrams in such a way that the 
propagator used in the perturbative expansion becomes that of Gaussian
approximation, and the zeroth order term of the effective action consists of 
the Gaussian one.

In the last section we have considered the quantum mechanical
anharmonic oscillator,
and have shown that the perturbative correction greatly
improves the Gaussian
approximation even at the first non-trivial ($\l^2$ contribution)
correction level.

One can easily compute the $\l^2$ contributions
to the effective potentials for the 
(1+1) and (2+1) dimensional $\f^4$ theories, and show that they agree 
with those of Cea and Tedesco\cite{cea}.
Our method appears to be simpler than the variational perturbation theory 
in practical applications and can also be used
in the cases of time-dependent systems.
Our method can easily be generalized to the cases of fermionic
and gauge field theories.

\newpage
\begin{center}
{\Large \bf Acknowledgements}
\end{center}
This work was supported in part by Korea Science and Engineering
Foundation under Grant 
No.95-0701-04-01-3 and 065-0200-001-2,
the Center for Theoretical Physics (SNU), and 
by the Basic Science Research Institute Program, Ministy of Education,
under project No. BSRI-96-2425.

\newpage

\newpage
\noindent Figure 1. Feynman diagrams for Eq.\ ({\protect \ref{j3}}).
\newline
\noindent Figure 2. Feynman diagrams for Eq.\ ({\protect \ref{j4}}).
\newline                    
\noindent Figure 3. Digrams contributing to the $\l^2$-order corrections to the
                    Gaussian effective action.
\noindent Table 1. ground state energy for anharmonic oscillator.

\newpage
\begin{table}
\caption{ground state energy for anharmonic oscillator \label{ground}}
\begin{tabular}{|c|c|c|c|c|c|c|c|c|}
 & \multicolumn{2}{c|}{1-loop} & \multicolumn{2}{c|}{2-loop} &
\multicolumn{2}{c|}{Gaussian} & \multicolumn{2}{c|}{improved Gaussian} \\  
\cline{2-9}
$\l$&
$E_0$ & error(\%) & $E_0$ & error(\%) &
$E_0$ & error(\%) & $E_0$ & error(\%)  \\  
\tableline
0 & 
0.5 & 0  & 0.5 &0  & 
0.5 & 0  & 0.5 &0   \\
0.24 &
0.5&      -1.43&        0.508&    0.048 &
0.507&   0.006&     0.507&  -0.0003  \\
2.4&
0.5&-10.6&        0.575& 2.84 &
0.560&0.21& 0.559&-0.037 \\ 
24 &
0.5&-37.8&        1.25&55.5&
0.813&1.09&     0.801&-0.37  \\
240 &
0.5&-66.8&       8.&432&
1.53&1.75&     1.49&-0.68 \\
2400& 
0.5&-84.0&       75.5&2311 &
3.19&1.95&   3.11&-0.79 \\
$\infty$ & 
 $\cdot$ & $\cdot$ & $\cdot$ & $\cdot$ &
$0.236 \! \lambda^\frac{1}{3}$ & 2.01& $ 0.230
\! \lambda^\frac{1}{3}$ & -0.82 
\end{tabular}
\end{table}


\begin{references}
\bibitem{barns} T. Barnes  and G.  I. Ghandour,  
                \prd{22}{924}{1980};
                P. M. Stevenson, 
                \prd{32}{1389}{1985};
                S.-Y.Pi and M.Samiullah,
                \prd{36}{3128}{1987};
                O.Eboli, R.Jackiw and S.-Y.Pi,
                \prd{37}{3557}{1988};
                S.K. Kim, W. Namgung, K.S. Soh and J.H. Yee,
                \prd{41}{3792}{1990};\prd{43}{2046}{1991};
                \ap{214}{142}{1992};
                S.K. Kim, J. Yang, K.S. Soh and J.H. Yee,
                \prd{40}{2647}{1989};
                S. Hyun, G.H. Lee and J.H. Yee,
                \prd{50}{6542}{1994};
                H.-J. Lee, K. Na and J.H. Yee,
                \prd{51}{3125}{1995};
                H.-J. Lee, S. Lee and J.H. Yee, 
                \jkps{29}{5751}{1996} and
                references therein.
\bibitem{polley} L. Polley and D.E.L. Pottinger, eds., 
               {\em Variational calculations in quantum field theory},
               World Scientific, Singapore (1987);
                L. Polley and U. Ritschel, 
                  \plb{221}{44}{1989};
                U. Ritschel,
                   \plb{227}{251}{1989}; \zp{47}{457}{1990}; 
                H. Verschelde and M. Coppens,
                   \plb{287}{133}{1992}.

\bibitem{cea}  A.Okopi\'{n}ska, 
               \prd{35}{1835}{1987};
               \prd{38}{2507}{1988};
                Ann. Phys.(N.Y.) {\bf 228}, 19 (1993).
              I. Stancu and P. M. Stevenson, 
                 \prd{42}{2710}{1990};
              P. Cea,
                \plb{236}{191}{1990};
              I. Stancu, 
                 \prd{43}{1283}{1991};
              P.Cea and L.Tedesco,
                 \plb{335}{423}{1994};\prd{55}{4967}{1997}.

\bibitem{cklee} B.S. DeWitt,
                  \prs{162}{1195}{1967};
                G.t' Hooft,
                in Acta Universitatis Wratislavensis no. 38,
                12th Winter School of Theoretical physics in Karpacz;
                Functional and probabilistic methods
                in quantum field theory vol. I (1975);
                L.F.Abbott,
                   \npb{189}{185}{1981};
                C. Lee,
                  \npb{207}{157}{1982};
                W.Ditrich and M. Reuter,
                  {\em Selected Topics in Gauge Theories},
                   Springer-Verlag (1985).

\bibitem{steve1} P. M. Stevenson, 
                \prd{30}{1712}{1984}.
\end{references}
\end{document}